# Molecular Dynamics of Polymer-lipids in Solution from Supervised Machine Learning

James Andrews[1,2], Olga Gkountouna[2], Estela Blaisten-Barojas[1,2,*]

*[1]Center for Simulation and Modeling, George Mason University,*
*[2]Department of Computational and Data Sciences, George Mason University,*
*Fairfax, Virginia 22030, USA.*

[*] Correspondence: blaisten@gmu.edu

February 24, 2022

## ABSTRACT

Currently machine learning techniques including neural networks are popular tools for materials and chemical scientists with applications that may provide viable alternative methods in the analysis of structure and energetics of systems ranging from crystals to biomolecules. However, efforts are less abundant for prediction of kinetics and dynamics. Here we explore the ability of three well established recurrent neural network architectures for forecasting the energetics of a macromolecular polymer-lipid aggregate solvated in ethyl acetate at ambient conditions. The solvated 4-macromolecule aggregate is considered to be a pre-micellar formation of finite life span. Data models generated from three recurrent neural networks, ERNN, LSTM and GRU, are trained and tested on nanoseconds-long time series of the intra-macromolecules potential energy and their interaction energy with the solvent generated from Molecular Dynamics and containing half million points. Our exhaustive analyses convey that the three recurrent neural network architectures investigated generate data models with limited capability of reproducing the energetic fluctuations and yielding short or long term energetics forecasts with underlying distribution of points inconsistent with the input series distributions. We propose an *in silico* experimental protocol consisting on forming an ensemble of artificial network models trained on an ensemble of series with additional features from time series containing pre-clustered time patterns of the original series. The forecast process improves by predicting a band of forecasted time series with a spread of values consistent with the molecular dynamics energy fluctuations span. However, the distribution of points from the band of forecasts is not optimal. Although the three inspected recurrent neural networks were unable of generating single models that reproduce the actual fluctuations of the inspected molecular system energies in thermal equilibrium at the nanosecond scale, the proposed *in silico* protocol provides useful estimates of the molecular fate.

## 1. Introduction

Machine learning techniques in materials science, chemistry, and physics has seen an impressive popularity growth over the past decade[1] for a multitude of applications[2-4]. Recurrent neural networks (RNN) were developed for sequence prediction. They differ from other neural networks in that they contain a hidden layer that operates on an ordered sequence where each step includes a hidden state that is updated as a function of its respective input features and the hidden state of the previous step[5,6]. The process entails use of a window of data from a sequence and predicting the following sequence data points moving forward one data point at a time. Under this perspective, RNNs are self-supervised learning approaches[7]. Most popular RNN architectures comprise the Elman RNN (ERNN)[8], the long-short term memory (LSTM)[9], and the gated recurrent unit (GRU)[10]. The ERNN is among the simplest RNNs containing one hyperbolic tangent function as the activation function. LSTM is a newer architecture explicitly designed to avoid the long-term dependency problem where temporally distant data have a vanishing gradient, or are "forgotten" by the network. The GRUs are modifications of LSTM containing fewer parameters while improving performance on certain tasks[11]. Recent alternative perspectives suggested combining a statistical optimized technique for projecting high dimensional data onto spaces of lower dimension to promote deep learning as applied to biomolecular function recognition[12]. LSTM has been used for calibrating Brownian particles force fields[13], for learning the constitutive laws of viscosity in history-dependent materials[14]. Supervised neural network learning with LSTM and GRU has addressed complex systems such as multiscale approaches linking molecular dynamics trajectories with continuum dynamics[15], to categorize the hydrogen bond mobility in water[16], for following the geometric changes of the SARS-CoV-2 spike protein in aqueous solutions[17], among others. There is evidence that LSTM and GRU have potential to be considered as alternative models to *ab initio* molecular dynamics[18,19] by predicting short time series forecasts given a fair amount of time series data on the atomic positions and velocities.

The use of polymers with amphiphilic behavior has increased the accessibility of drug delivery in nanotherapeu-tics[20-22]. Macromolecules such as lipids, surfactants, and copolymers have the ability of aggregating into diverse structures in aqueous solutions and are capable of morphing into other structures when the solvent condition changes[23]. Such structural characteristics entail both, the interaction between the macromolecules composing the aggregates and the extent the solution affects the macromolecular structure. Ultimately, the main factors controlling the equilibrium structures formed are thermodynamics and the solution modulated intra-aggregate forces. These aggregate structures are fluid-like, soft, with the macromolecules coiling, twisting, rotating, or diffusing within each aggregate due to thermal motions[24]. Thus, soft structures do not exhibit a definite shape or size and are characterized by distributions about average values. Time is also important since property averages require a sufficiently long time evolution for avoiding biases from conformational local minima. From the perspective of computational simulations, to follow the fate of these solvated macromolecules is a challenge because emulation of actual experimental conditions require very large systems at the atomic scale and very lengthy computer simulations at the nanosecond scale and beyond. This is the case for biomolecules[25] but also for other macromolecules including polymers. Related to polymers in particular, LSTM networks have been used for drug discovery[26], polymeric materials development for solar cell applications[27], benchmarking neural network predictions on enumeration of polymer sequence space[28], predicting signal peptides and their cleavage sites from protein sequences[29], engineering dielectric parameters based on nonlinear structure of polymers[30], among others.

We investigate in this article the prospect of employing RNN long term forecasts for estimating the future behavior of energetic properties of a liquid solution containing a solute aggregate of polymer-lipid macromolecules in an organic solvent at ambient conditions. Our goal is to assess with an *in silico* experimental protocol how good and useful are the artificially generated energetic forecasts of a polymer system at the nano spatial and temporal scales in the context of recent molecular dynamics predictions concerning that particular polymer-lipid assembly[31]. First, with extensive all-atom molecular dynamics (MD) the system containing an aggregated cluster of four polymer-lipid macromolecules (1808 atoms) solvated in ethyl acetate (224,000 atoms) is equilibrated at 300 K and 101.325 kPa. Worth mentioning is the literature scarcity on describing the temporal fate of small solute aggregates formed with a few macromolecules in solution while the full system is in equilibrium[32]. Second, we continue the simulation and save to disk both, the atomic positions and the solute aggregate and solvent energy properties, along the MD trajectories. The energy time series are to be used for the learning tasks, training and testing, of the RNN data models. The dynamics of polymers is slow requiring tens to hundreds of nanoseconds to observe the evolution of processes such as the life span of a macromolecular solute in a solvent. In fact, it is not excluded that the macromolecular aggregate will dissociate over time if ethyl acetate proves to be a good solvent for the polymer-lipid material considered in our work[33]. However, the dissociation process may require long and costly MD simulations. Hence, exploration of alternative approaches for estimating even a portion of the system future evolution is the main motivation of this paper. Does a RNN forecast of the aggregate energy properties over a time period of 20% the time length of series entering in the RNN model training/testing properly describe the aggregate future fate? Our inspection for answering this question employs the RNN cyber implementation of pyTorch[34], one of the top three preferred, thorough, with solid foundations, well recognized, and accessible deep learning frameworks able of accelerating the path from research prototyping such as the one we describe in this paper to production deployment when the prototype is proven to be useful.

Our *in silico* prototype experiment is based on applying RNN data models to forecast the time evolution of two system energies without having to predict the temporal behavior of the atomic positions and velocities that serve to calculate them. If successful, such prototype protocol will require MD simulations of a few nanoseconds and, yet, would permit access to forecasted system energies without extending the simulation time. Additionally, the protocol eliminates the need of storing the atomic positions required for calculating the forecasted energetics, thus saving on terabytes of archival storage and substantial electric power for the storage upkeep. Therefore, from the ERNN, LSTM, and GRU implementation, we generated data models trained and tested on the time series of two energy properties of the polymer-lipid aggregate in the solvent. These models not only yield an approximation of the input series but additionally serve for predicting short (0.5 to 5 ps) and long time forecasts (1 ns). We found that these three RNN models trained on one or more energy series tend to forecast smooth time series around the average value of the series used for training/testing. Such outcome is short of capturing the distribution of points underlying the forecasted energetics. Based on this evidence, we augment our prototype protocol by building an ensemble of training time series that yield an ensemble of data models that generate and ensemble of energetics forecasts. When inspected as a whole the ensemble of forecasts is perceived as a band of values that span the range of the MD energy fluctuations. However, the distribution of the ensemble of forecasted energy values differ from the distributions of the original time series.

# 2. Models and Methods

## 2.1 Molecular Description

DSPE-PEG(2000) or 1,2-Distearoyl-sn-glycero-3-phosphoethanolamine-poly(ethylene glycol)$_{45}$amine, is a block polymer-lipid, with chemical formula $(C_2H_4O)_{45}C_{44}H_{86}N_2O_{10}P$, and containing 452 atoms. Multiple applications of this macromolecule include thermo-sensitive liposomal nanoparticles and in the formation of micelles, disks, vesicles, and bilayers that are commonly assembled for therapeutic drug delivery[35-39]. We term this macromolecule DSPE-PEG. The organic solvent ethyl acetate (EA), $C_4H_8O_2$, is often used for the fabrication of nanoparticles[20,40,41]. The system under study here was composed of an aggregate formed by four DSPE-PEG macromolecules solvated by 16,000 EA molecules yielding a solution with 0.787% by mass solute relative concentration. Both EA and DSPE-PEG were modeled using the all-atom generalized Amber force field (GAFF)[42,43] with custom calculated restrained electrostatic potential atomic charges[31,33,44], which was combined with the compatible Amber-Lipid17[45] force field for the DSPE portion of the polymer-lipid macromolecule. The GAFF-Lipid17 were transferred to the GROMACS 2018-2020[46-48] package used for our MD simulations.

## 2.2 Molecular Dynamics Approach

Three system samples were prepared for evaluating our prototype protocol, as is a general practice in wet lab and *in silico* experiments. Hence, three different macromolecular cluster geometries were generated from 1 ns MD simulations in vacuum at 300 K along which the macromolecules aggregated into a flexible globular-like cluster. Each of these aggregated assemblies was solvated in EA and brought to thermodynamics equilibrium through 10 ns NPT-MD runs using the Parrinello-Rahman[49,50] and Berendsen[51] pressure coupling, 1 fs time step, periodic boundary conditions, 1.4 nm cutoff, and particle-mesh electrostatics as implemented in the GROMACS 2018-2020[46-48]. The three systems equilibrated at a density of $906.3\pm0.1$ kg/m$^3$ at 300 K, comparable to the density of pure EA[44]. These three simulations are herein termed Sets 1, 2, and 3. At the conclusion of the three equilibrations, the structure of the solvated macromolecular aggregate in each set differed significantly between them with positional root mean squared atomic deviation (RMSD) of 1.74 nm between Set 1 and Set 2, 1.74 nm between Set 2 and Set 3, and 1.58 nm between Set 1 and Set 3. The RMSDs were calculated with the VMD[53] corresponding plugin[54] that aligns the compared structures by optimizing consecutive rotations between specified groups of atoms.

After the equilibration process, the three sets underwent 10 ns NVT-MD runs at a temperature of 300 K using the velocity rescale approach[54]. A buffer of the first 5 ns simulation time was not used in the forthcoming analyses, while the subsequent 5 ns of trajectories were retained, with the full system configuration saved every 10 fs for a total of 500,000 configurations for each set under study. With the saved trajectories, the desired energetics was calculated including the total system potential energy $E_{total}$, the potential energy of the full solvent $E_{solvent}$, the four macromolecules intra-potential energy sum PE, the four macromolecule-solvent interaction energies IE, and the macromolecular cluster cohesive energy $E_{coh}$. Both, IE and $E_{coh}$ were sums of the Coulomb and Lennard-Jones terms between solvent atoms and DSPE-PEG atoms for the former and between atoms in different DSPE-PEG macromolecules for the latter. Calculation of the energetics entailed repeated use of the GROMACS *rerun* feature[46], which involved preparation of separate files with the atomic coordinates of each system subcomponent for which the potential energy required calculation. The $E_{solvent}$ had fluctuations on the order of 0.03% and its contribution to the $E_{total}$ was greater than 99.99%. The system total energy, the solvent energy, and their difference $\Delta E = E_{total} - E_{solvent}$ were considered basically constant for the purpose of the forthcoming RNN learning analysis that focused on energy components entering in $\Delta E$:

$$\Delta E = PE + IE + E_{coh} \qquad (1)$$

The energy time series with 500,000 points for each energy and set considered were collected from the MD simulations. Table 1 lists the series means and standard deviation, evidencing that the PE and IE magnitudes were more than one order of magnitude larger than $E_{coh}$ and displayed smaller standard deviations. Hence, the analyses focused primarily on PE and IE.

Table 1: Mean and standard deviation values of the energies entering in Eq. 1 along the 5 ns MD trajectories for the Sets 1, 2, 3. Also provided are the corresponding values for the last 1 ns interval from the 4th to the 5th ns evolution. Energies are in MJ/mol.

|       | time (ns) | $\Delta E$      | $E_{coh}$        | PE              | IE               |
|-------|-----------|-----------------|------------------|-----------------|------------------|
| Set 1 | 0-5       | $2.22 \pm 0.17$ | $-0.52 \pm 0.08$ | $8.97 \pm 0.15$ | $-6.23 \pm 0.15$ |
|       | 4-5       | $2.29 \pm 0.17$ | $-0.54 \pm 0.06$ | $8.98 \pm 0.16$ | $-6.15 \pm 0.16$ |
| Set 2 | 0-5       | $2.20 \pm 0.17$ | $-0.47 \pm 0.05$ | $8.77 \pm 0.17$ | $-6.11 \pm 0.16$ |
|       | 4-5       | $2.13 \pm 0.18$ | $-0.46 \pm 0.03$ | $8.76 \pm 0.15$ | $-6.16 \pm 0.14$ |
| Set 3 | 0-5       | $2.28 \pm 0.18$ | $-0.71 \pm 0.08$ | $9.09 \pm 0.16$ | $-6.10 \pm 0.20$ |
|       | 4-5       | $2.31 \pm 0.19$ | $-0.76 \pm 0.04$ | $9.05 \pm 0.15$ | $-5.98 \pm 0.18$ |

The data preparation addressed the creation of two ensembles of time series containing ten participants each. In Ensemble_{100} each 5 ns time series having 500,000 time points was sampled 10 times with systematic sampling, with the initial ten consecutive time points of the series as starting points and populating the sampled series with points selected after a fixed 100 fs sampling interval. This probability sampling method yielded a manifold of 10 time series, each of them spanning the full 5 ns time series and containing 50,000 time points separated by 100 fs. In Ensemble_{10}, the original series of 500,000 time points were segmented into 10 consecutive series, each one of them with a span of 0.5 ns with time intervals of 10 fs.

For the forthcoming analyses, the data in series belonging to both ensembles were normalized by subtracting the series average to each time point and dividing by the standard deviation.

## 2.3 Machine Learning Approach

The first machine learning (ML) analysis consisted in identifying groups of time points in the PE and IE energy time series. These patterns served to characterize the data from a machine learning perspective. Clustering is a ML unsupervised learning technique. We employed the expectation maximization (EM)[55] clustering algorithm as implemented in scikit-learn[56,57], a Python-based library of machine learning methods. The EM features (or descriptors) used were the ten PE and ten IE time series of Ensemble_{100} for a total of twenty features and 50,000 instances. The clustering process was performed on each of the Set 1-3, independently. The number of retained clusters was based on the sum of squares among clusters $SSA = \sum_{i=1}^{k} n_i (<y_i> - <y>)^2$, with $<y_i>$ being the mean of clusters containing $n_i$ points and $<y>$ the mean of all of the points to be clustered. When $SSA$ reached a plateau value with at most a 2% change, we identified six relevant clusters, termed cluster 0-5. The task yielded six time patters embedded in each time series to be used later within the neural network (NN) analyses. These clusters were considered as distinct themes that could modulate the NN models. Thus, when employing them for the NN training, the task was referred to as *cluster seeding* of the pristine energy series.

The second ML analysis was based on evaluating the possibility of extending the simulated PE and IE time series into the future without performing additional simulations. A subfield of ML is deep learning and its foundational methodology of artificial NNs. In particular, RNNs are recognized as the best ML methods for predicting how sequenced data may be cyber-continued without using the technique employed for generating the original sequence.

We based our ML *in silico* experiment on ERNN, LSTM and GRU, three very well established and extensively used RNNs available in a variety of cyber libraries and software frameworks. Our computing implementation required multiple scripts that were written in Python 3.6. The RNN functions were implemented as included in the PyTorch 1.7.0 package[34,58].

RNNs are a class of NNs that allow previous processed output to be used as input to the next step while keeping hidden states. Figure 1 is a schematic representation of how a basic recurrent neuron operates. The blue cells are the one step at a time feeds from the PE and IE series points to the network orange cells where an *activation* process combines the current with previous processed input, gives an approximated output depicted yellow, and simultaneously passes the previously processed feed to the next orange activation cell and to a hidden layer. The process is done one step at a time across the input time window. The neuron activation cells are functions that modulate a weighted mixing of the previously processed step with the pristine input at the next step and are different for the ERNN, LSTM and GRU. The recurrent neurons are organized as belonging to a layer. At each time step one neuron activates an input that is sent to all other neurons in the layer and the compounded activation is propagated one-step at a time, simultaneously, by all neurons in the layer.

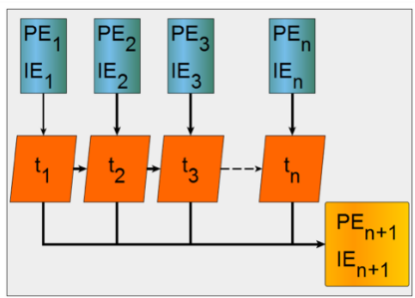

Figure 1: Schematic workflow of a basic recurrent neuron. Input is in blue, neuron activation is in orange, output occurs step by step depicted with a line that generates sequentially the approximated series points in yellow.

Hyperparameters for the RNNs were carefully determined after testing a multitude of different alternatives. Taking together the PE and IE time series as *input*} yielded the best results. The *time window* size was set to 50 consecutive time points for each series. The metrics for quality assessment of each created NN data model was based on training on a *fold* containing the input time series first 80% points and testing on a second fold containing the remaining 20% points. Hence, each generated NN data model had its own errors. The *loss* and *validation error* are the errors incurred on the training and testing regions of the series, respectively. The loss was calculated as the mean squared error MSE between each of the N targeted series points $y_i$ and the N corresponding $x_i$ NN approximated points:

$$MSE = \frac{1}{N}\sum_{i=1}^{N}(x_i - y_i)^2 \qquad (2)$$

The MSE is minimized with respect to the NN parameters in an iterative manner, each iteration is termed *epoch*. The root mean squared error is termed RMSE.

Concerning the RNN user-adjustable hyperparameters, *neurons* (or nodes) and *layers* refer to the width and depth of the network, respectively. Each layer contains a number of neurons that pass their output to the following layer. *Dropout* refers to an additional layer where each element has a probability of being set to zero, which improves regularization and prevents co-adaptation of neurons[59]. Meanwhile, the *regularization factor* is a constant applied to weights during training that helps preventing overfitting and the *learning rate* is the optimizer rate of approaching the minimum. Our optimal RNN hyperparameters for series in Ensemble$_{100}$ were: 300 neurons, one layer, zero% dropout, zero regularization factor, and $10^{-4}$ learning rate. Figure 2 summarizes the behavior of these hyperparameters depicting the validation errors for the LSTM and GRU NNs when one parameter was varied while the others were kept constant. We observed that considering additional layers beyond the first or a high regularization factor hindered significantly the network ability of learning from the data. Additionally, we observed that the number of neurons affected the rate by which the data model approached the minimum error, hence, neurons primarily affected the number of epochs needed for training rather than the ability for the network to learn. The NN data model ability to produce a small loss was more affected by the size and granularity of the training data than by the selection of the hyperparameters. In fact, as visually evidenced in Fig. 2, variation of the hyperparameters yielded averages within the standard deviation of each other, suggesting that the choice of these hyperparameters are relatively inconsequential.

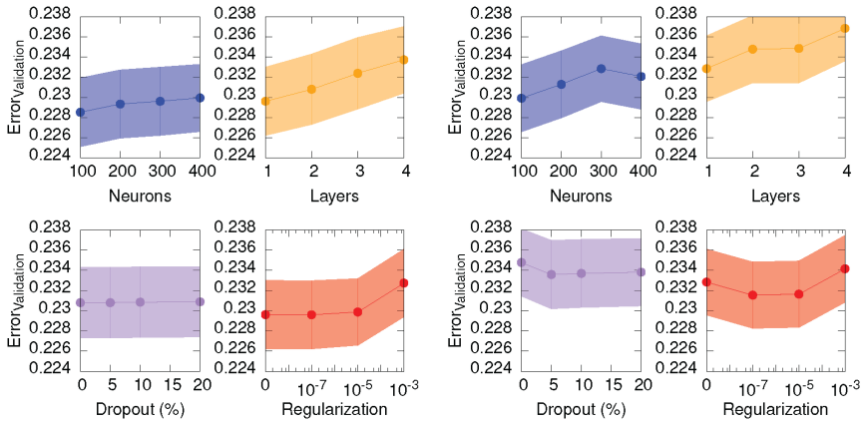

Figure 2: Comparison of the validation error for various hyperparameter values entering the GRU NN on the left and LSTM on the right. Data was obtained for Ensemble$_{100}$ Set 3. Errors are shown after 40 epochs, or at the minimum validation error if sooner as in the case of 4 Layers. Dots are mean values over the ensemble of series and shaded area is $\pm$ one standard deviation.

Concerning the forecasting ability of the NNs, we adopted the average over the testing folds known as *evaluation on a rolling forecasting origin*[60] as evaluation of the short term forecast accuracy. For this process the cross-validation were folds containing the full series minus one point out of the last $m$ points for testing, with $m$ =1-50 limited by the size of one time window. In fact, the origin at which the forecast was based rolled forward in time when started after the $m^{th}$ point. We considered data models predicting a single time step forward. Hence, for predicting more than one time step the output is appended to the input time window and the oldest time point is dropped. The outcome was a sliding time window that feeds into the next prediction. Consequently, errors in previous predicted values propagated to the subsequent predictions and compounded. Short term forecasts were done for the two energies, PE and IE. When considering the full series with 500,000 time points, the short term predictions covered only 0.5 ps, while considering series members of the Ensemble$_{100}$ allowed for 5 ps short term forecasting. Challenging the validity of these forecasts, we continued the process to generate long term forecasts extending from the last time window up to 1 ns forecast by appending the predicted time points to the end of the series as to propagate the sliding input window forward in time, one step at a time. Our goal was to forecast a period of time equivalent to a fifth of the original time series length. Hence long term forecasts were 1 ns in length. The best model was expected to maintain the lowest error for the longest time, however, along the long term forecasts there were no known data to determine the MSE.

## 3. Results

Recapitulating, the MD simulation gave rise to the 5 ns trajectories along which the energy times series of interest correspond to PE, total intra-macromolecule potential energy, IE, the total interaction of the macromolecular cluster with the solvent, and E$_{coh}$, the cohesive energy that keeps the aggregate of four DSPE-PEG macromolecules together. These series are visualized in Fig. 3 for the three simulations, Sets 1, 2, and 3.

The distribution of the 500,000 points comprising the PE and IE energy series is depicted in Fig. 4. The PE and IE energy series distribution corroborated the expected gaussian distribution of MD potential energies in equilibrium. Meanwhile, the IE distribution was indicative that there was energy exchange between the solvent and the macromolecular aggregate. Indeed, lower IE corresponded to higher E$_{coh}$ and vice-versa. Eventually, with enough time, we expected that the aggregate would dissociate into its four macromolecules with E$_{coh}$ tending to zero.

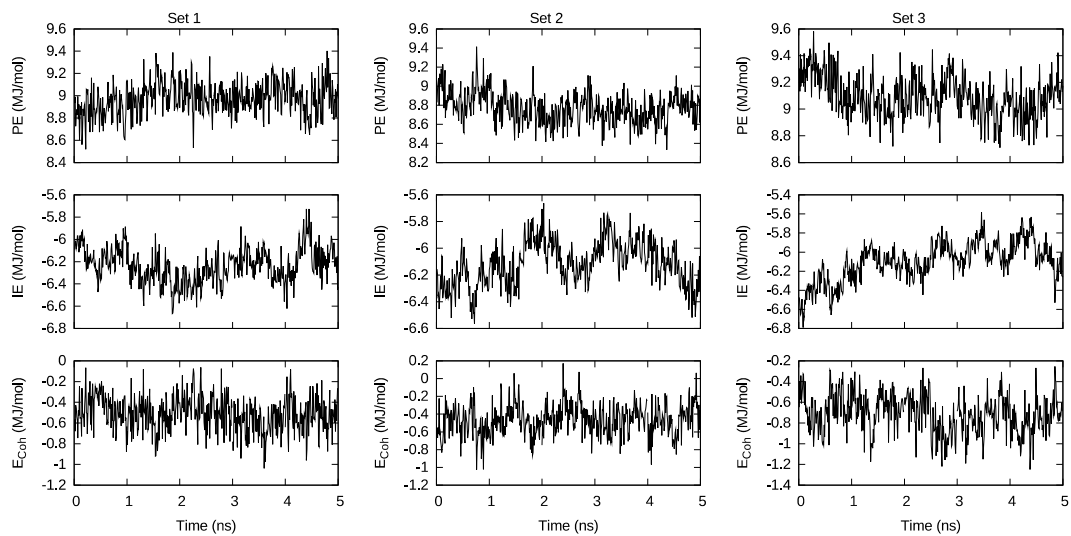

Figure 3: Visualization of the total intra-macromolecule potential energy PE, the total interaction energy between macromolecules and solvent molecules, and the cohesive energy that keeps the macromolecular cluster together $E_{coh}$. Time series are provided for Sets 1, 2, 3

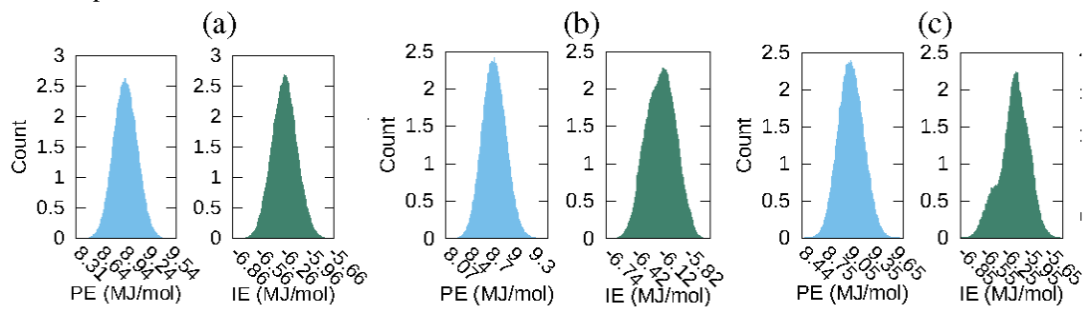

Figure 4: Distribution of the time values in the PE and IE time series shown in Fig. 3. (a) Set 1, (b) Set 2, (c) Set 3.

The split of the two energy properties PE and IE into six EM data clusters for the Ensemble$_{100}$ evidenced that the ten series members of the ensemble of each energy type behave very similar to one another. In addition, the clustering outlook of the three studied sets was comparable. The data distribution associated to each of the ten member series of the ensemble was even and unimodal as shown in Fig. 5 for set 1. It is evident that the IE data were most influential in producing the clustering since the PE datasets displayed only a gentle increase in value over the six data clusters. In addition, the violin plots feature a very similar density estimation of the underlying distribution for the ten participating time series in Ensemble$_{100}$. Table 2 provides the percentage of time points in each cluster, the energetics average values and standard deviations. The macromolecular aggregate radius of gyration $R_g$ in Set 2 was slightly larger than in the other two sets, peculiarity that persisted in the split datasets generated by the EM clustering process.

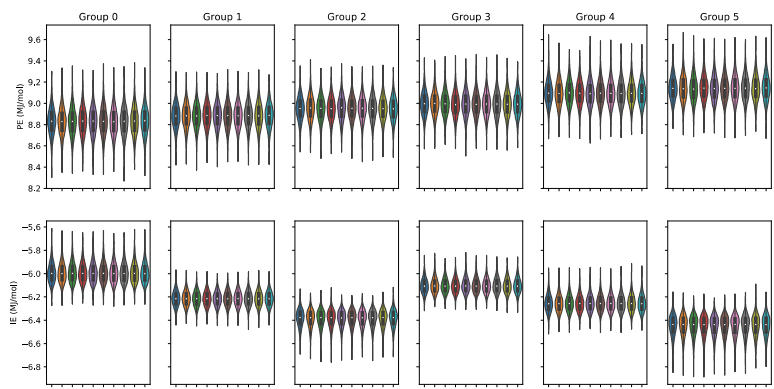

Figure 5: PE and IE violin plots of the Ensemble$_{100}$ series clustered with EM into six data clusters for Set 1.

Table 2: Energetic properties of EM clustered groups from the time series in Ensemble100. EM groups are numbered in ascending order of PE. The radius of gyration $R_g$ of the DSPE-PEG aggregate is mass weighted

| | Cluster No. | 0 | 1 | 2 | 3 | 4 | 5 |
|---|---|---|---|---|---|---|---|
| | % of samples | 11.7 | 24.4 | 16.0 | 16.9 | 20.4 | 10.6 |
| Set 1 | PE (MJ/mol) | $8.83 \pm 0.12$ | $8.88 \pm 0.11$ | $8.95 \pm 0.11$ | $8.99 \pm 0.11$ | $9.10 \pm 0.11$ | $9.14 \pm 0.11$ |
| | IE (MJ/mol) | $-5.99 \pm 0.08$ | $-6.215 \pm 0.06$ | $-6.385 \pm 0.07$ | $-6.10 \pm 0.06$ | $-6.25 \pm 0.07$ | $-6.45 \pm 0.09$ |
| | $E_{coh}$ (MJ/mol) | $-0.53 \pm 0.08$ | $-0.508 \pm 0.08$ | $-0.49 \pm 0.07$ | $-0.553 \pm 0.09$ | $-0.547 \pm 0.08$ | $-0.516 \pm 0.08$ |
| | Rg (nm) | $2.05 \pm 0.08$ | $2.08 \pm 0.08$ | $2.14 \pm 0.08$ | $2.06 \pm 0.08$ | $2.09 \pm 0.09$ | $2.14 \pm 0.10$ |
| | % of samples | 16.9 | 17.8 | 19.8 | 21.8 | 12.6 | 11.1 |
| Set 2 | PE (MJ/mol) | $8.64 \pm 0.10$ | $8.65 \pm 0.14$ | $8.77 \pm 0.11$ | $8.80 \pm 0.11$ | $8.94 \pm 0.12$ | $8.96 \pm 0.12$ |
| | IE (MJ/mol) | $-6.07 \pm 0.07$ | $-5.882 \pm 0.07$ | $-6.241 \pm 0.07$ | $-6.042 \pm 0.07$ | $-6.357 \pm 0.08$ | $-6.17 \pm 0.07$ |
| | $E_{coh}$ (MJ/mol) | $-0.45 \pm 0.05$ | $-0.47 \pm 0.05$ | $-0.45 \pm 0.05$ | $-0.46 \pm 0.05$ | $-0.49 \pm 0.06$ | $-0.49 \pm 0.07$ |
| | Rg (nm) | $2.28 \pm 0.10$ | $2.29 \pm 0.12$ | $2.28 \pm 0.10$ | $2.28 \pm 0.10$ | $2.26 \pm 0.08$ | $2.29 \pm 0.10$ |
| | % of samples | 13.1 | 16.4 | 19.7 | 21.2 | 14.0 | 15.6 |
| Set 3 | PE (MJ/mol) | $8.93 \pm 0.12$ | $8.95 \pm 0.11$ | $9.06 \pm 0.11$ | $9.12 \pm 0.11$ | $9.24 \pm 0.11$ | $9.25 \pm 0.13$ |
| | IE (MJ/mol) | $-5.83 \pm 0.09$ | $-6.06 \pm 0.07$ | $-6.18 \pm 0.09$ | $-5.95 \pm 0.09$ | $-6.14 \pm 0.09$ | $-6.43 \pm 0.10$ |
| | $E_{coh}$ (MJ/mol) | $-0.76 \pm 0.08$ | $-0.68 \pm 0.07$ | $-0.67 \pm 0.07$ | $-0.75 \pm 0.06$ | $-0.73 \pm 0.07$ | $-0.66 \pm 0.08$ |
| | Rg (nm) | $2.11 \pm 0.07$ | $2.08 \pm 0.09$ | $2.05 \pm 0.09$ | $2.08 \pm 0.08$ | $2.26 \pm 0.07$ | $1.99 \pm 0.08$ |

Concerning the generation of data models from the three RNN architectures, Fig. 6 illustrates the training loss and testing validation error resulting for time series in Ensemble$_{10}$ (left) and Ensemble$_{100}$ (right). The time series in both ensembles have equal number of points. However, the training loss is significantly higher when the interval between time points is larger, as occurs in Ensemble$_{100}$. The figure also evidenced that for the two ensembles the GRU achieved the lowest loss at the end of training in all cases. The ERNN was the most sensitive to the training process in the early epochs, containing the widest range of error. The speed at which the networks converge to a minimum error relates to the complexity of the neuron. ERNN neurons containing a single activation function reach the minimum relatively quickly, although this NN is the least descriptive of the data as evidenced by its comparatively poor validation performance. The most complex neuron, LSTM, displays the slowest training process from among the three architectures, outperforming ERNN and achieving comparable results than GRU. We concluded that for our system the GRU model outperformed the other NNs, as evidenced by the lowest validation error of Fig. 6.

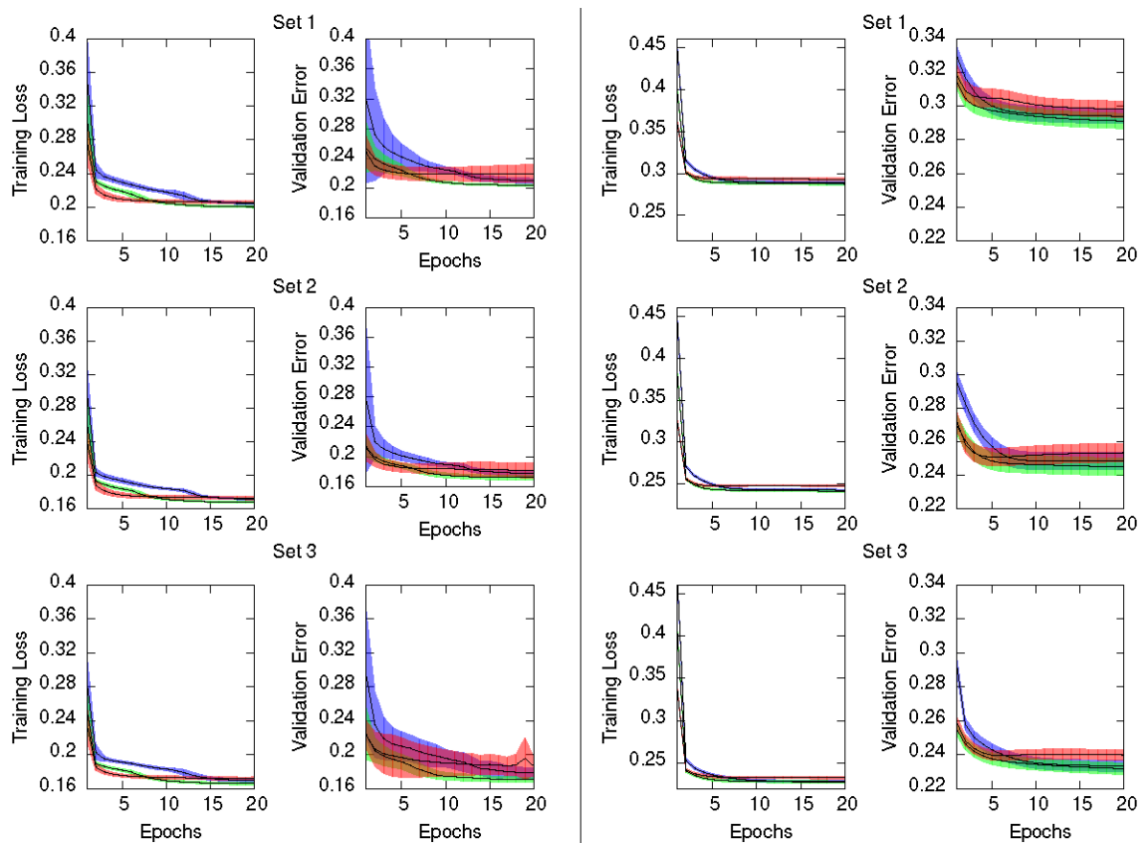

Figure 6: Mean training loss and testing validation error for the approximated series generated from the ERNN (red), LSTM (blue), and GRU (green) data models. Series in Ensemble$_{10}$ (left) and Ensemble$_{100}$ (right). Shaded regions denote $\pm$ one standard deviation from the mean. Loss and validation errors are the MSE of Eq. 2 pertaining to the training and testing regions of each series in the ensemble, respectively.

Forthcoming results on the GRU time series forecasting constituted our best output scenario. Ten GRU models generated for series in Ensemble$_{100}$, considering 4.995 ns of the PE and IE time series as the training data and using the last 0.005 ns for starting short terms forecasts and yet having time points to evaluate the error incurred. Then, each model propagated its prediction along 50 iterations that spanned 0.005 ns. Moreover, for each member of Ensemble$_{100}$, six additional GRU models were generated in which the clustered PE and IE series obtained before were added to the model training for a total of 4 participating series. This task produced a set of sixty additional cluster-seeded models for which a forecast over 0.005 ns was also obtained. This process was repeated for data in Sets 1, 2, and 3. Results from these short term forecasts are illustrated in Fig. 7, where dark and light colored points identify the GRU model forecasts without and with cluster-seeded input, respectively. The RMSE of each of the 70 model predictions spanned from about 0.1 MJ/mol for the cluster-unseeded model predictions to almost 0.2 MJ/mol for the cluster-seeded models, as seen from the histograms of RMSE in Fig. 7. This figure also depicts the distribution of predicted points from the 70 inspected models clearly indicating that the GRU-predicted time series distributions from the cluster-seeded models approximated fairly the PE distribution shape of the original series of Fig. 4 with a maximum-minimum spread of about half the standard deviation reported in Table 1. Meanwhile, resemblance between actual and GRU-generated IE distributions fades dramatically in shape and spread.

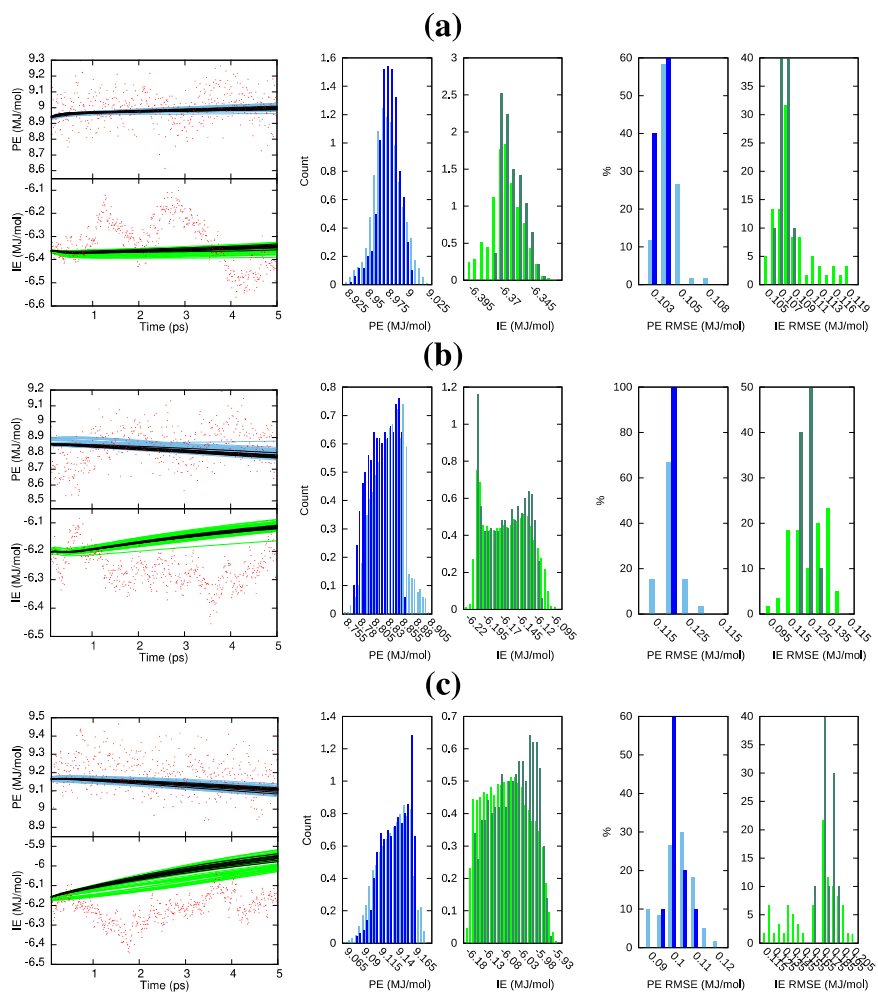

Figure 7: Short time forecast predictions and the corresponding RMSE over 0.005 ns from models trained over 4.995 ns. Plot (a) corresponds to Set 1, (b) to Set 2, and (c) to Set 3. Left panels are the forecasts. Middle panels are the distributions of predicted values. Right panels are the RMSE of predicted points as a percentage of the number of models. Dark colors (blue, green) correspond to 10 GRU models that are not cluster-seeded. Light colors (blue, green) identify the 60 GRU models that were cluster-seeded. Red dots are the true values from the MD simulation.

One ns period was our desirable time span for the energetics long term forecast, thus, the long term forecasting ability of the GRU models created from Ensemble$_{100}$ was also explored. The seventy GRU models trained on 4.995 ns for the Ensemble$_{100}$ series obtained previously, without and with cluster-seeding, were reutilized as initial GRU models for the process of predicting the future 1 ns behavior of PE and IE time series. The last time window of each series in the ensemble was used to define different initial times for the long term forecasts.

Figure 8 depicts the forecast of PE in blue and IE in green along the targeted time span of 1 ns into the future. Dark-colored and light-colored dots in the figure identify predictions without and with cluster-seeded models and the shaded area illustrates the maximum-minimum dispersion of predictions coming from the 70 models and the 50 initial steps for the forecast. In a nutshell, the future predicted values tend to gather around specific energy regions maintaining a fairly constant spread around them.

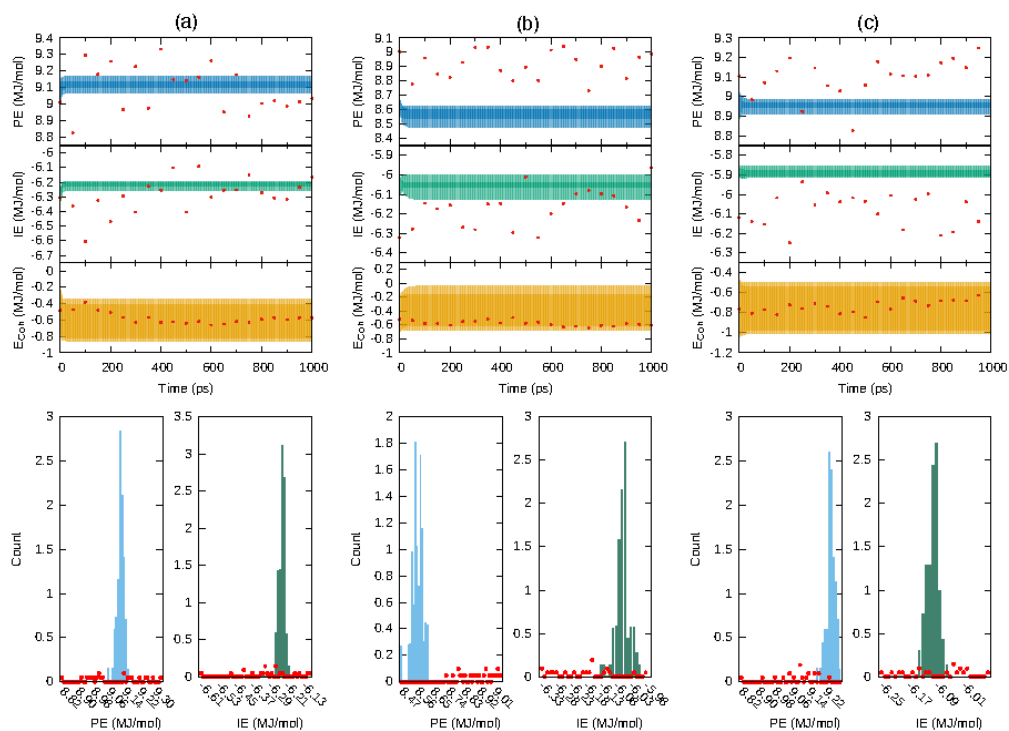

Figure 8: Forecast prediction of PE and IE for a time span of 1 ns for series in Ensemble100 (top) and the distribution of predicted PE and IE values (bottom). The E$_{coh}$ is estimated from the forecasted energetics, Eq. 1, and $\Delta$E mean during the 4-5 ns MD trajectory time reported in Table 1. (a) Set 1, (b) Set 2 and (c) Set 3. The dark shaded region denotes forecasted energies from GRU models without cluster-seeding. Light shaded area denotes forecast energies from GRU models with cluster-seeding. Normalization of the histograms is with respect to 70x50,000 evaluated time points. Red dots are true energies from a posterior MD simulation for each set.

We went back and continued the three simulations for 1 ns saving configurations every 0.05 ns. These new instantaneous simulation points are depicted as red points in Fig. 8. During the span of this forecast we assess that the GRU models are able of predicting the forecast behavior on average, however, being unable of generating the characteristic thermodynamic fluctuations that are revealed in the MD simulations. Averages of the 70 predicted forecasts along 1 ns were 9.13, 8.57, 8.95 MJ/mol for PE and -6.23, -6.05, -5.88 MJ/mol for IE of Set 1, 2, 3, respectively. These averages are 0.1 - 0.2 MJ/mol either larger or smaller than the actual mean over 1 ns of Table 1. Summarizing, the GRU architecture generated data models that forecasted the energetics with series displaying a distribution of points that agglomerated close to the mean with an overall spread smaller than the energetics standard deviations reported in Table 1. The short and long time energetics forecasts of Figs 7,8 are markedly different from the thermodynamic fluctuations of the potential energies under study and are not able of maintaining their statistics. Indeed, the studied energies are gaussian-distributed or superposition of gaussian functions as expected for systems in thermodynamic equilibrium and evidenced in Fig. 4; this statistical property appears to be lost even by building the ensemble of forecasts as visualized in the middle panels of Fig. 8.

## 4. Discussion

Our GRU results show that the data models obtained were instrumental in providing a band of energetics forecasts around the mean of the fed energy series involved in the training/testing. However, the ensemble of energetics forecasts displayed a distribution of energy values atypical of the thermodynamics fluctuations characteristic of the molecular dynamics energies despite the small losses and validation errors of the GRU data models reported in Fig. 6. This is indicative that the ground truth energy fluctuations are not accessed by well trained GRU model. We have proven that the time series granularity does not improve this deficiency, despite the more than 30% loss decrease when the series time granularity is reduced by a factor of 10, as seen in Fig. 6. A peculiarity of the GRU data model forecast of energy values is a distinctive early convergence toward a particular value whereas, thereafter, the forecasted values became almost constant as function of time. In order to remediate this shortcoming we have shown that considering an ensemble of GRU data models enables a maximum-minimum spread of predicted energy forecasts of about the standard deviation of the actual MD calculated energies. However, the spread of the forecasted energies retains strongly the character acquired in the short term forecast, which in turn depends on energy values close to the end of the time series employed during training/testing.

Relatively, the short term forecast RMSEs were very similar between PE and IE as shown in Fig. 7. Including the clustering information into the GRU data model increases the spread of predicted forecasts as evidenced in Figs. 7,8. We demonstrated that the temporal behavior of the $E_{coh}$ energy derived from the PE and IE forecast values constituted a reasonable estimate of the macromolecular aggregate fate along 1 ns within a 95% confidence interval from the MD values, as evidenced in Fig. 8. Polymers have a slow dynamics as evidenced by the PE and IE time autocorrelation functions depicted in Fig. 9. These autocorrelation functions showed that the correlation time of the IE is on the order of 1 ns, indicating that the system memory would persist over the time span of the attempted forecasts in what concerns how the solvent affects the macromolecular aggregate. On the other hand, the relatively short correlation time of the PE is characteristic of the macromolecular internal vibrational modes that display a sharp drop in the autocorrelation function along a few picoseconds time. It is disappointing that GRU generated time series forecasts predicted narrowly confined fluctuations while the PE and IE time series possess distinct and different correlation times and, when considered together, these energy series have proven optimal in building the GRU data model with the lowest error.

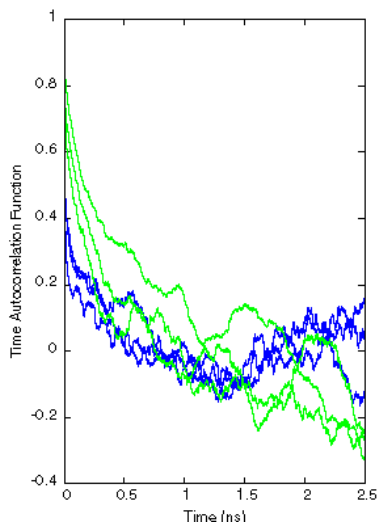

Figure 9: Time autocorrelation function of the PE (blue) and IE (green) as a function of the time lag for Sets 1, 2, 3.

There are challenges unique to recurrent neural networks and deep learning methods when applied to molecular modeling and simulation that we observed at the nano spatial and temporal scale that differentiate this *in silico* protocol from applications of such techniques to other systems. For example, inductive biases associated with molecular modeling are very different from the traditional and proven uses of RNNs such as natural language processing[61]. Indeed, molecular models are invariant to positional translations, invariant to atom permutation of identical particles, and continuous regarding changes in position, which are all absent to RNN inductive biases. The complexity and size of the molecular system do contribute to these challenges, the more data is furnished, the less useful the NN computer implementations become. The learning objectives involved in molecular modeling are very different from traditional deep learning tasks. Machine learning techniques excel at classification problems, but their use to solve regression and extrapolation problems that are common in molecular modeling remains challenging due to the lack of target functions smoothness.

Assessment of a RNN model ability to forecast is difficult, where the uncertainty of the predictions is unknown. To help measure uncertainty of the model, we generated an ensemble of predictions and examined the differences between the different predictions. However, such preparation was computationally expensive and demanded human participation. Furthermore, the trained models are particular to the system under study. There is no expectation that such models would be transferable to other molecular systems by changing the values of a few parameters, limiting their practicability. Systems simulated with molecular dynamics are ergodic, which ensures that the time-to-time potential and interaction energies should maintain a non-regular temporal behavior. Our work shows that the current RNNs and available framework implementation did not handle effectively the fluctuations associated with thermodynamic energy properties in equilibrium of our system, a paradigm at the nano temporal and spatial scales.

# 5. Conclusions

The ERNN, LSTM, and GRU neural networks were trained on the potential energies time series of a nanoscale DSPE-PEG aggregate solvated in ethyl acetate obtained from MD in the nanosecond time regime. The system was complex, had 225, 808 atoms, and the analyzed energetic properties were series with half a million time points. The main goal was to predict energy forecasts that would extend in time the MD simulations by $10^6$ MD time steps. The targeted energetics of the solvated DSPE-PEG aggregate included the intra-macromolecules potential energy and the macromolecules-solvent interaction energy. The resulting RNN models were extensively trained and tested evidencing 1% training and validation errors, hence, ensuring that the original time series and their underlying statistical distribution of points were reproduced well. We demonstrated that for the system studied here, the energies forecasts predicted well the mean values of the desired energies in the short term. Nonetheless, the data models failed at maintaining the underlying statistics and energy values spread during the long term forecast period. In fact the forecast consisted in a set of points with a very narrow spread, resembling a delta function. A second goal of our research was to provide a reliable estimate of the macromolecular aggregate lifetime and fate during the forecasted time span. We demonstrated that the aggregate cohesive energy calculated from the two forecasted energies yielded an excellent estimate, predicting that the macromolecular aggregate persisted associated during the time span of the forecast. By using time series of system energies, rather than the time series of atomic positions, we demonstrated the feasibility of generating artificial RNNs models for forecasting the temporal behavior of a nanoscale polymer-lipid system energetics. To alleviate the RNN difficulty of maintaining the proper underlying statistical distribution of the forecasted time points, a machine learning protocol was proposed encompassing two strategies. The first strategy consisted in generating an ensemble of shorter time series from the original ones, all of them covering the time span of the original series. The second strategy consisted in identifying a group of time patterns from the original time series based on machine learning clustering and seeding each of the ensemble series with one of these patterns. The combination of both strategies yielded an ensemble of 70 time series of each of the two desired energies. Independent RNN models were generated for each seeded tine series in the ensemble, and these RNN models produced an ensemble of forecasts of the desired energies. The ensemble of forecasts yielded a band of predicted energies with a spread of about half the expected one.

This work demonstrates with an *in silico* experiment that for systems at the nano temporal and spatial scales the traditional RNN architectures, which have proven their effectiveness in sequence regimes such as natural language processing and music generation, may require further tuning from experts in artificial learning for becoming proficient when addressing time series from systems similar to the one described in this work.

## Acknowledgments

We acknowledge partial support by the Commonwealth of Virginia, USA, under the 4-VA 2018-2021 grants "Scalable Molecular Dynamics" and "Is AI capable of identifying meaningful patterns in the temporal behavior of solvated macromolecules?." Computations were done in the two supercomputer clusters of the Office for Research Computing of George Mason University, USA.